# Generation, interaction and reduction of filaments as a nonlinear process.

L. M. Kovachev[a], D.A. Georgieva[b], N. Nedyalkov[a]
[a]Institute of Electronics, Bulgarian Academy of Sciences
72 Tzarigradsco shoussee, 1784 Sofia, Bulgaria
[b]Faculty of Applied Mathematics and Computer Science, Technical University of Sofia, 8 Kliment Ohridski Blvd., 1000 Sofia, Bulgaria;
e-mail:lubomirkovach@yahoo.com

**Abstract:** The absence of ionization and observation of white continuum in the initial moment of filamentation of powerful femtosecond laser pulses, propagating in silica glasses, as well as the filamentation without plasma channels observed in the experiments in air, forced us to look for other nonlinear mechanisms of description the above mentioned effects. For this reason we present in this paper new parametric conversion mechanism for asymmetric spectrum broadening of femtosecond laser pulses towards the higher frequencies in isotropic media. This mechanism includes cascade generation with THz spectral shift for solids and GHz spectral delay for gases. The process works simultaneously with the four-photon parametric wave mixing. The proposed theoretical model gives very good coincidence with the experimental data. In addition we demonstrate that the nonlinear model describes the process of reduction of the number of filaments trough two mechanisms of nonlinear interaction: attraction due to cross-phase modulation and energy exchange due to four-photon parametric wave mixing.

## 1. Introduction

The established in 1985 technology of amplification through chirping of laser pulses gave the opportunity of fast progress and construction of femtosecond (fs) laser facilities producing high intensity field of order of $10^{12} \div 10^{19}\ W/cm^2$. On the other hand the duration of the pulse was quickly reduced from picoseconds till 5-6 femtoseconds. This progress allowed the observation of new nonlinear effects relied to multi-photon ionization, high harmonics emission, plasma defocusing, tunnel ionization and others. In the experiments investigating the propagation of fs pulses in air, in gas medium and quartz glass was observed also other nonlinear effects such as filamentation [1, 2], GHz and THz emission [3, 4], rotation of the polarization plane [5], merging and energy exchange between filaments [6, 7]. At the first experiments in air on filamentation [1], the typical experimental set up produces nonlinear focus by self-focusing, where the field intensity reaches $10^{14} - 10^{15}\ W/cm^2$. This intensity is high enough for plasma generation and observation of the effect of plasma defocusing. Therefore the first theoretical models [8, 9] relate the waveguide propagation with a balance between self-focusing and plasma defocusing. Even nowadays in leading journals are being published papers containing the standard model where to the equations of paraxial nonlinear optics are added terms containing tunnel and multi-photon ionization, higher orders of nonlinearities ($\chi^{(5)}, \chi^{(7)}, ...$), Raman effect and others. Generally, the problem is not solvable analytically and the idea is that the equations should be solved numerically using very powerful computers. Throughout parameters variation the aim is to be obtained relatively stable waveguide propagation of the filament. Until now this approach has not demonstrated soliton-like propagation at a distance longer than 1-2 meters, while on a vertical trace the real single filament in the experiment reaches till about ten kilometers. *There are deep contradictions between plasma kind of interpretation of filamentation and the real experiments*. For example, in experiments with long focus distance lenses, with purpose to avoid the nonlinear focus initially in [10, 11] and then in almost all leading laboratories is observed filamentation without ionization of the medium. Another fundamental contradiction between experiment and the standard theory is that the measured intensity in the stable filament outside the nonlinear focus is of order of $10^{11} - 10^{12}\ W/cm^2$, which is two to three orders lower than the intensity needed for defocusing by ionization. In all experiments on filamentation the effect is being observed if the power of the laser pulse is a bit higher than the critical one for self- focusing $5-10\quad GW$. The critical power value for self focusing is $P_{cr} = \pi(0.61\lambda_0)^2/(8n_0n_2)$. It defines also the intensity of the field for laser pulses. For pulses having spot diameter of $100-200\ \mu m$ the intensity is $I \approx 10^{12}\ W/cm^2$. Recently were obtained single broadband filaments in quartz glass [12]. The absence of ionization in this process leads to the occurrence of the following basic questions:

1) *What kinds of equations describe the diffraction, the dispersion and the nonlinear propagation of broad-band (attosecond and phase-modulated femtosecond) pulses?*
2) *What is the physical process that leads to asymmetrical spectrum broadening which is observed experimentally in direction from infrared region to the visible region?*
3) *What kinds of mechanisms are the ones of merging and energy exchange between the filaments?*

## 2. Limits of applicability of the amplitude approximation in the linear and nonlinear optics.

As already is discussed in the literature [13, 14, 15], the nonlinear propagation in isotropic materials of ultra-short optical pulses is described generally by vector integro-differential nonlinear wave equation

$$\Delta\vec{E} - \frac{1}{c^2}\frac{\partial^2\vec{E}}{\partial t^2} = \frac{4\pi}{c^2}\frac{\partial^2}{\partial t^2}\left\{\begin{array}{l}\int\limits_0^\infty R^{(1)}(\tau)\vec{E}(t-\tau)d\tau + \\ \int\limits_0^\infty\int\limits_0^\infty\int\limits_0^\infty R^{(3)}(\tau_1,\tau_2,\tau_3)\left[\vec{E}(t-\tau_1,r)\cdot\vec{E}(t-\tau_2,r)\right]\vec{E}(t-\tau_3,r)d\tau_1 d\tau_2 d\tau_3\end{array}\right\}. \quad (1)$$

In scalar approximation the Taylor expansion of the square of the linear $k_{lin}^2(\omega) = \omega^2\varepsilon(\omega)/c^2$ and nonlinear $k_{nl}^2(\omega) = 4\pi\omega^2\chi^{(3)}(\omega)/c^2$ wave-numbers leads to amplitude representation of the wave equation, usually approximated to the second order of the linear dispersion and the zero order of the nonlinear dispersion. This approximation was called Slowly Varying Amplitude Approximation (SVAA), which assumed that the equation is valid for narrow-band pulses with significant number of cycles under the envelope. It is not difficult to show, that using the series expansion of the square of the linear wave-number $k^2(\omega)$ near the carrying frequency of the laser pulses $\omega_0$, and using the presentation of the electrical field of the kind of $E(x,y,z,t) = A(x,y,z,t)\exp[i(k_0 z - \omega_0 t)]$, the linear integro-differential polarization operator of the wave equation (1) takes the form:

$$\frac{4\pi}{c^2}\frac{\partial^2 P_{lin}(r,t)}{\partial t^2}=\frac{4\pi}{c^2}\frac{\partial^2}{\partial t^2}\left(\int_0^\infty R^{(1)}(\tau)E(t-\tau)d\tau\right)=$$

$$-\left(k_0^2 A+\left(\frac{\partial k^2}{\partial\omega}\right)_{\omega_0}\frac{\partial A}{\partial t}+\frac{1}{2}\left(\frac{\partial^2 k^2}{\partial\omega^2}\right)_{\omega_0}\frac{\partial^2 A}{\partial t^2}+\sum_{m=3}^{\infty}\frac{1}{m!}\left(\frac{\partial^m k^2}{\partial\omega^m}\right)_{\omega_0}\frac{\partial^m A}{\partial t^m}\right)\exp(i(k_0 z-\omega_0 t)). \qquad (2)$$

Such an expansion is correct if the differential operator series in the brackets of the right-hand side of the equation is strongly convergent. To make a quantitative analysis of the series convergence, we normalize in time the amplitude function and its derivatives in (2). The normalized amplitude of a localized in time pulse can be written as:

$$A=A_0 A'(x,y,z,t/t_0) \qquad (3)$$

where $t_0$ denotes the initial temporal duration of the pulse, $A'\to 0$ if $t\to\pm\infty$, and $\max A'=1$. Substituting (3) in the Taylor series (2), we obtain a normalized function series of the linear polarization operator:

$$\frac{1}{c^2}\frac{\partial^2 P_{lin}(r,t)}{\partial t^2}=-A_0\left(k_0^2 A'+\left(\frac{\partial k^2}{\partial\omega}\right)_{\omega_0}\frac{1}{t_0}\frac{\partial A'}{\partial t}+\frac{1}{2}\left(\frac{\partial^2 k^2}{\partial\omega^2}\right)_{\omega_0}\frac{1}{t_0^2}\frac{\partial^2 A'}{\partial t^2}+\sum_{m=3}^{\infty}\frac{1}{m!}\left(\frac{\partial^m k^2}{\partial\omega^m}\right)_{\omega_0}\frac{1}{t_0^m}\frac{\partial^m A'}{\partial t^m}\right)\times \qquad (4)$$

$$\times\exp(i(k_0 z-\omega_0 t))$$

The normalization makes the *functional series* (4) as series of the derivatives of the normalized localized function (distribution)

$$f'(A')=A',\ \frac{\partial A'}{\partial t},\ \frac{\partial^2 A'}{\partial t^2},\dots\ \frac{\partial^m A'}{\partial t^m}\propto f'(A')^0,\ f'(A')^1,\ f'(A')^2\dots.f'(A')^m.., \qquad (5)$$

where with upper indexes are signed the order of derivatives. On the other hand, the normalized *numerical Taylor series* of the wave-number square takes the form:

$$g'(k^2)=k_0^2;\ \left(\frac{\partial k_0^2}{\partial\omega}\right)_{\omega_0}\frac{1}{t_0};\ \frac{1}{2}\left(\frac{\partial^2 k_0^2}{\partial\omega^2}\right)_{\omega_0}\frac{1}{t_0^2};\ \frac{1}{3!}\left(\frac{\partial^3 k_0^2}{\partial\omega^3}\right)_{\omega_0}\frac{1}{t_0^3};\dots,\frac{1}{m!}\left(\frac{\partial^m k^2}{\partial\omega^m}\right)_{\omega_0}\frac{1}{t_0^m},.. \qquad (6)$$

$$=g'(k^2)^0;\ \ g'(k^2)^1;\ \ g'(k^2)^2\dots.g'(k^2)^m...,$$

where the superscripts denote the corresponding derivatives. The numerical Taylor series of the wave-number square expansion around the carrier frequency (6) turns into a series, for which the dimensionality of each term is equal to the dimensionality of the first term – the square of the wave-number ($cm^{-2}$). If expressed in terms of functional (5) and numerical (6) series, the right-hand side of Eq.(4) becomes

$$R(A,k^2)=A_0\left(g'^0 f'^0+g'^1 f'^1+g'^2 f'^2+\dots.g'^m f'^m\dots\dots\right). \qquad (7).$$

The maximal value of the function $f'^0$ in the first term of the function series (7) is 1, and the maximal values of the derivatives $f'^m$ are always less than one. This is a typical property of the normalized distribution functions. Each term of the series (7) is a product of a numerical Taylor series (6) and a normalized, *majorant* series of a distribution function and its derivatives. For the series (7) we choose the *maximal values* of the function terms in (5). Thus, the series (7) turns into a numerical series of the type

$$P=A_0\left(g'^0+g'^1/2+g'^2/2^2+\dots.g'^m/2^m\dots\dots\right)$$

For this reason, a sufficient condition for the convergence of the series of the linear polarization operator (4) is the convergence of the numerical series

$$P_m\left(k^2,t_0\right)=\left(\frac{1}{m!}\left(\frac{\partial^m k^2}{\partial\omega^m}\right)_{\omega_0}\frac{1}{t_0^m}\frac{1}{2^m}\right)\quad m=0..\infty. \tag{8}$$

The series $P_m(k^2,t_0)$ will be studied for optical pulses of wavelength $\lambda = 800\,nm$ propagating in air and having various temporal duration $t_0$. More specifically, we will express the different pulse half-widths $t_0$ in terms of the number of oscillations of the carrier frequency at a level $e^{-1}$ from the pulse maximum $t_0 = nT_0$; $T_0 = n_0(800nm)\lambda^{800nm}/c$. For a wavelength $\lambda = 800\,nm$ we obtain $T_0 \cong 2.6\times10^{-15}$ sec. The Ciddor formula [16] has been applied to calculate the dielectric constant $\varepsilon(\omega)$, the square of the wave-number and its derivatives in (8). We investigate the series (8), varying N from N = 100 (the case of slowly varying amplitudes) to N ~ 1 (one optical cycle pulse) for the dispersion expansion up to the 5-th order. Let us use the following notation for the *m*-th component of the Taylor series:

$$\beta_0 = k_0^2;\quad \beta_m = P_m = \left(\frac{1}{m!}\left(\frac{\partial^m k^2}{\partial\omega^m}\right)_{\omega_0}\frac{1}{t_0^m}\frac{1}{2^m}\right). \tag{9}$$

The values of the numerical expansion (9) for a spectrally-narrow pulse (266 fs, $N = 100$) present a strong convergence of the series. The value $\beta_3/\beta_2$, i.e. the ratio of the third to second series term in (9) is of the order of $10^{-7}$. For this reason, truncation of the expansion (2) at the second order of the dispersion is sufficient for describing the dynamics of a narrow-band wave-packet propagating in air.

Let's now investigate the convergence of the series (9) for broad-band pulses. Table 1 presents the results for the series truncated to the 5-th order terms for a spectrally-broad pulse with two cycles within the amplitude function $N = t_0/T_0 = 2$ ($t_0 = 5.3\,fs$).

| $\beta_0$ | $\beta_1$ | $\beta_2$ | $\beta_3$ | $\beta_4$ | $\beta_5$ |
|---|---|---|---|---|---|
| $6.17X10^9$ | $4.9X10^8$ | $9.7X10^6$ | 7.9 | $9.01X10^{-2}$ | $2.8X10^{-4}$ |

**Table 1** *Values of m-th component of the Taylor series (9) for spectrally-broad 5.3 fs pulse ($N = 2$). The series is also strongly convergent. The ratio of the third to the second order of dispersion $\beta_3/\beta_2$ is of the order of $10^{-6}$.*

The estimation shows that for gas medium with weak dispersion, truncation of the series (4) at the second order is sufficient for the description of pulses with even one or two optical cycles under the envelope.

In addition we perform such calculation for fused silica, where the dispersion $\varepsilon(\omega)$ was calculated using the Sellmeier dispersion formula [17]. The series is also strongly convergent, but the ratio $\beta_3/\beta_2$ of the third to the second dispersion term is now $10^{-3}$. As it can be seen, for fused silica it is also possible to truncate the series (4) to the second order, but the third order dispersion can play a role as a small parameter, even in spectral region where $\beta_2 \neq 0$. The analysis of the series expansion of the linear polarization (2) or (4) leads to the following conclusion:

***It is possible to use amplitude approximation to describe the dynamics of broad-band optical pulses with duration down to one optical cycle, using the second-order amplitude approximation in air, and including the third dispersion order as a small parameter in fussed silica***.

**3. The nonlinear operator of cubic type $\vec{P}_{nl} = n_2\left(\vec{E}\cdot\vec{E}\right)\vec{E}$**

We generalise the Maker and Terhune's nonlinear polarization operator $\vec{P}_{nl} = n_2\left[A\left(\vec{E}\cdot\vec{E}^*\right)\vec{E} + B\left(\vec{E}\cdot\vec{E}\right)\vec{E}^*/2\right]$, to nonlinear operator, which in addition includes the generation of third harmonics. In this case the nonlinear polarization in isotropic materials can be written as

$$\vec{P}_{nl} = n_2\left(\vec{E}\cdot\vec{E}\right)\vec{E} \ . \qquad (10)$$

We will derive vector amplitude equations for two component electrical field

$$\vec{E} = \frac{\left(A_x \exp\left[ik_0\left(z - v_{ph}t\right)\right] + c.c\right)}{2}\vec{x} + \frac{\left(A_y \exp\left[ik_0\left(z - v_{ph}t\right)\right] + c.c\right)}{2}\vec{y}, \qquad (11)$$

where $k_0$ is the carrying wave number, $v_{ph}$ is the phase velocity, $A_x = A_x(x,y,z,t)$ and $A_y = A_y(x,y,z,t)$ are the components of the vector amplitude of the electrical field $\vec{A} = \left(A_x, A_y, 0\right)$ and $\omega_0 = k_0 v_{ph}$ is the carrying frequency of the electrical field. Using standard procedures, the corresponding System of Vector Nonlinear Amplitude Equations (SVNAE) up to second order of linear dispersion can be written as

$$\begin{aligned}
&-2ik_0\left[\frac{\partial A_x}{\partial z} + \frac{1}{v_{gr}}\frac{\partial A_x}{\partial t}\right] = \Delta A_x - \frac{\left(k_0 v_{gr}^2 k''+1\right)}{v_{gr}^2}\frac{\partial^2 A_x}{\partial t^2} + \\
&k_0^2\tilde{n}_2\left[\frac{1}{3}\left(A_x^2 + A_y^2\right)A_x \exp\left(2ik_0\left(z - v_{ph}t\right)\right) + \left(\left|A_x\right|^2 + \frac{2}{3}\left|A_y\right|^2\right)A_x + \frac{1}{3}A_x^* A_y^2\right] \\
&-2ik_0\left[\frac{\partial A_y}{\partial z} + \frac{1}{v_{gr}}\frac{\partial A_y}{\partial t}\right] = \Delta A_y - \frac{\left(k_0 v_{gr}^2 k''+1\right)}{v_{gr}^2}\frac{\partial^2 A_y}{\partial t^2} + \\
&k_0^2\tilde{n}_2\left[\frac{1}{3}\left(A_x^2 + A_y^2\right)A_y \exp\left(2ik_0\left(z - v_{ph}t\right)\right) + \left(\left|A_y\right|^2 + \frac{2}{3}\left|A_x\right|^2\right)A_y + \frac{1}{3}A_y^* A_x^2\right]
\end{aligned}, \qquad (12)$$

where $\tilde{n}_2 = \frac{3}{8}n_2$, $v_{gr}$ is the group velocity, $\beta = k_0 v_{gr}^2 k_0''$ and $k_0''$ is the dispersion of the group velocity. In all coordinate systems, laboratory, local time $\left(t' = t - z/v_{gr}, z' = z\right)$ and Galilean $\left(z' = z - v_{gr}t, t' = t\right)$ the group velocity add additional phase to the nonlinear terms, associated with the Third Harmonics (TH) generation. In this way, instead on TH frequency, these terms generate on frequency, which is equal to three times of the Carrier-Envelope Frequency (3CEF) or $\omega_{nl} = 3k_0(v_{ph} - v_{gr})$. 3CEF is with GHz frequency delay in gases and with THz frequency shift in solids. This is clearly seen, if we rewrite the SVNAE in Galilean coordinate system

$$
\begin{aligned}
&-2i\frac{k_0}{v_{gr}}\frac{\partial A_x}{\partial t}=\Delta_\perp A_x-\frac{\beta+1}{v_{gr}^2}\left(\frac{\partial^2 A_x}{\partial t^2}-2v_{gr}\frac{\partial^2 A_x}{\partial t\,\partial z}\right)-\beta\frac{\partial^2 A_x}{\partial z^2}\\
&\quad+k_0^2\tilde{n}_2\left[\frac{1}{3}\left(A_x^2+A_y^2\right)A_x\exp\left(2ik_0\left(z-\left(v_{ph}-v_{gr}\right)t\right)\right)+\left(\left|A_x\right|^2+\frac{2}{3}\left|A_y\right|^2\right)A_x+\frac{1}{3}A_x^*A_y^2\right],\\
&-2i\frac{k_0}{v_{gr}}\frac{\partial A_y}{\partial t}=\Delta_\perp A_y-\frac{\beta+1}{v_{gr}^2}\left(\frac{\partial^2 A_y}{\partial t^2}-2v_{gr}\frac{\partial^2 A_y}{\partial t\partial z}\right)-\beta\frac{\partial^2 A_y}{\partial z^2}\\
&\quad+k_0^2\tilde{n}_2\left[\frac{1}{3}\left(A_x^2+A_y^2\right)A_y\exp\left(2ik_0\left(z-\left(v_{ph}-v_{gr}\right)t\right)\right)+\left(\left|A_y\right|^2+\frac{2}{3}\left|A_x\right|^2\right)A_y+\frac{1}{3}A_y^*A_x^2\right]
\end{aligned}
\quad . (13)
$$

This is the basic SVNAE which we apply to investigate the nonlinear regime of propagation of ultra-short optical pulses. The linear terms allow description of paraxial diffraction and dispersion for narrow-band pulses and non-paraxial, Fraunhofer type diffraction of attosecond and phase-modulated femtosecond pulses [18]. The front nonlinear terms leads to coherent GHz generation (93-94 GHz) in air. The nonlinear operator includes also self-action term, cross-phase modulation and degenerate four-photon parametric mixing terms between the optical components.

## 4. The initial moments of generation of a filament in BK7 glass

### 4.1 *The experiment*

The experimental setup for observation of the initial moment of filamentation and for generation of broadband spectrum from femtosecond pulses is presented on Fig. 1. It is composed from Ti-Sapphire femtosecond laser 1, reflected mirror 2, attenuator 3, focused lens 4, BK7 glass sample with 5 mm thickness 5. Behind the sample is situated spectrometer 6.

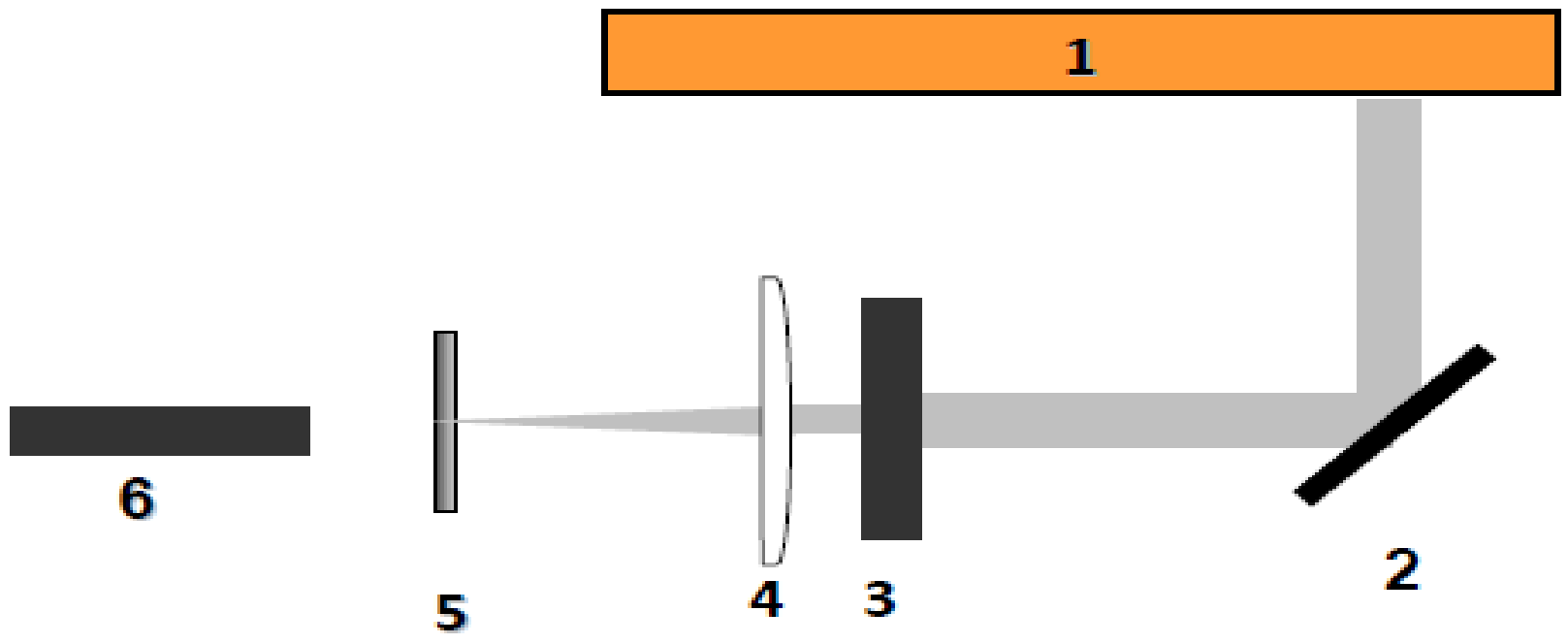

*Fig. 1. Experimental setup for observation of initial moment of filamentation in glasses. 1- Ti-Sapphire femtosecond laser; 2- reflected mirror 3- attenuator; 4- focused lens, 5- BK7 glass sample; 6- spectrometer.*

In the experiment 100 femtosecond pulse with wavelength 788 nm passed trough the BK7 glass plate with 5 mm thickness. The power of the pulse is slightly above the critical for self-focusing. Throughout spectrometer was measured the spectrum of the pulse in the region from 400 nm up to 950 nm. The result of measurements is presented on Fig. 2.

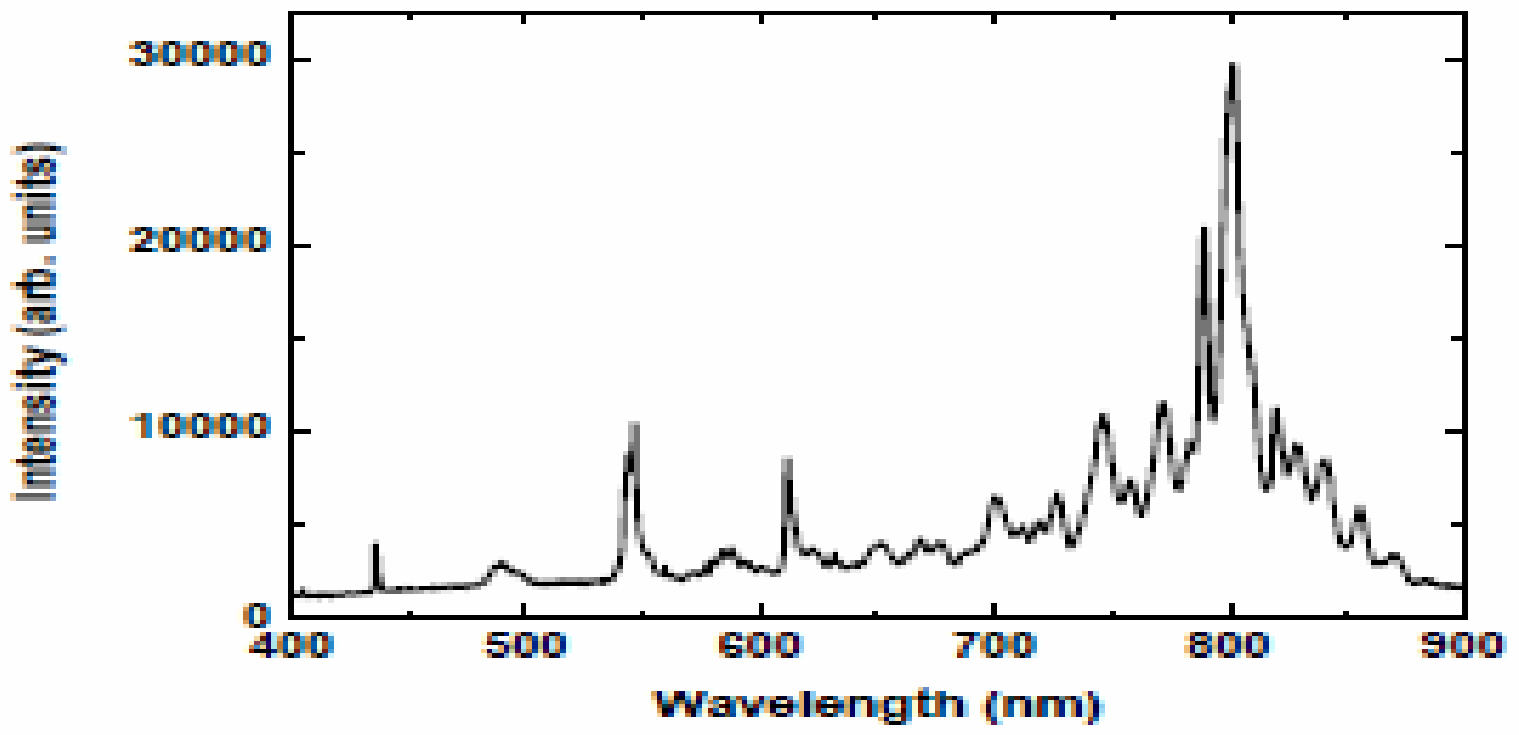

*Fig. 2. The spectrum of 100 fs pulse passed trough 5 mm BK7 glass sample. Spectral components from 450 nm up to 870 nm are clearly seen. The typical asymmetric conversion towards the short wavelengths shows that we observe the initial moment of filamentation process.*

In the spectrum are observed spectral components in the whole range from 450 nm up to 870 nm. For part of them conditions for Four-Photon Parametric mixing (FPPM) can be seen. It is well known that the FPPM processes are symmetrical in respect to main wavelength. Therefore appears the natural question: What kind of nonlinear process leads to asymmetrical broadening of the spectrum additionally to the FPPM?

4.2 *Physical model. Scalar approximation*

A careful analysis of the spectrum shows that the spectral distance between some of the components is with THz spectral shift. In the experiment the pulses propagate in glass with thickness 5 mm. That is why we use 1+1 dimensional scalar approximation of SVNAE (13) to calculate the spectral and spatial deformation of the laser pulse passing trough the glass (i.e. we neglect the diffraction). This restriction is due to the facts, that the dispersion and diffraction lengths are significant larger than the thickness of the sample. The scalar equation describing pulse propagation at these small distances becomes

$$-2i\frac{k_0}{v_{gr}}\frac{\partial A_x}{\partial t} = -\beta\frac{\partial^2 A_x}{\partial z^2} + k_0^2 n_2\left[\frac{1}{3}A^3{}_x \exp\left(2ik_0\left(z-\left(v_{ph}-v_{gr}\right)t\right)\right)+\left|A_x\right|^2 A_x\right], \qquad (14)$$

where $k_0$ is the carrying wave number, $v_{gr}$ is the group velocity, $v_{ph}$ is the phase velocity, $A_x$ is the amplitude envelope polarized linearly in $x$ direction, $\beta = k_0 v_{gr}^2 k'' << 1$ is the dimensionless dispersion parameter and $n_2$ is the nonlinear refractive index. It can be seen from Eq. 1, that instead on third harmonics $\omega_{nl} = 3k_0 v_{ph}$, the pulse generates on frequency delay $\omega_{nl} = 3k_0(v_{ph} - v_{gr})$ which is in GHz region for gases and THz region for solids [4]. For glasses the THz shift corresponds to wavelength shift $\Delta\lambda$ = 10 - 30 nm in respect to the pulse carrying wavelength ($\lambda_0$ = 800 nm) depending on the material of the glasses. The frequency emission falls in the spectrum wing of 100-200 fs pulse towards the shorter wavelengths (higher frequencies). This leads to asymmetrical spectrum broadening towards the shorter wavelengths. The amplified wavelength $\lambda_1 = \lambda_0 - \Delta\lambda$ of the optic wave is near the spectrum maximum, and if the intensity is high enough, becomes possible the occurrence of new emission again at spectral distance of $\Delta\lambda$ = 10 - 30 nm. As a result, a new wave will appear shifted at $2\Delta\lambda$. The joint action of this cascade process, along with the processes of four-photon mixing, would lead also to emission of weaker signals at wavelengths $\lambda_3=2\lambda_0-\lambda_1$ and $\lambda_4=2\lambda_0-\lambda_2$. Since $\lambda_3$ is shifted from $\lambda_0$ at same spectral distance $\Delta\lambda$=10-30 nm towards the longer wavelengths in respect to the main wavelength, it can act as basic wave and the wave $\lambda_0$ as signal wave in the process of THz emission. The result is that the energy transfer in the initial moment of emission will be again towards the shorter wavelengths. The above described mechanism of joint action of THz emission and FPPM processes could describe the asymmetrical ultra-broadening of the filament's spectrum from the infrared toward the visible region. To prove this physical model we need to use the dispersion $n(\lambda)$ of a BK7 glass from *www. refractiveindex.info* and to calculate first the group-phase velocity difference.

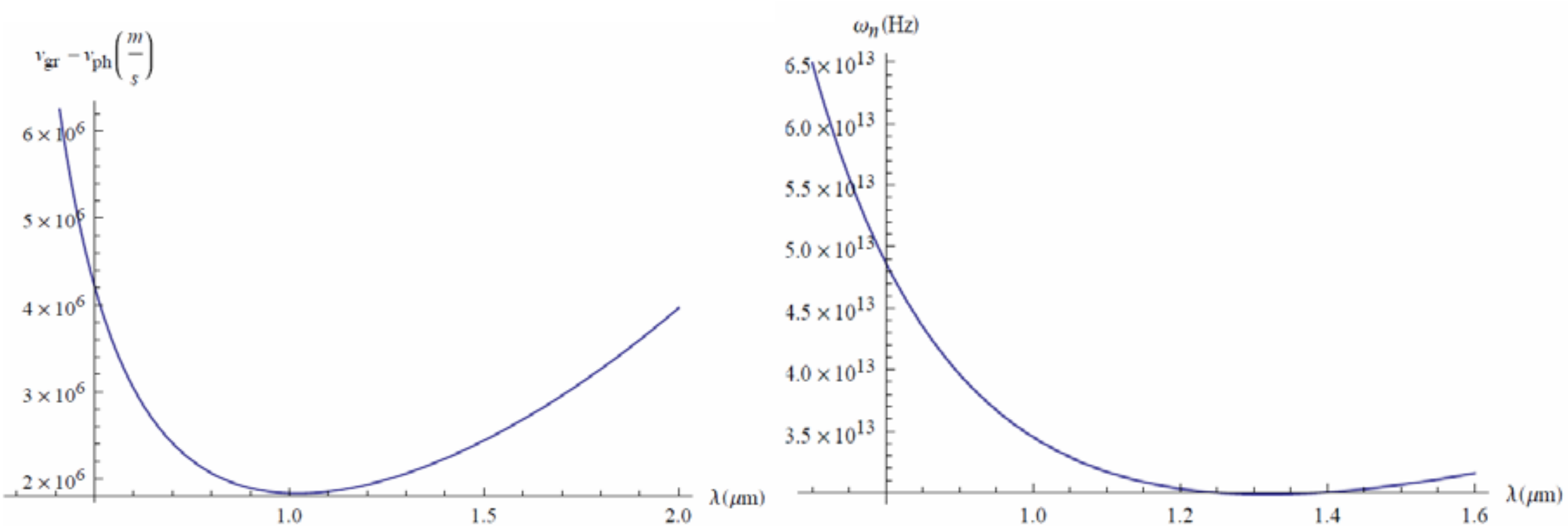

**a)** **b)**

*Fig 3. a) Group-phase velocity difference, calculated for BK7 glass in the region 780 nm up to 2 µm. b) The nonlinear frequency shift is* $\Delta\omega_{nl}^{bk7}=3\omega_{cep}=3k_0(v_{ph}-v_{gr})$. *As it can be seen (Fig. 3b) the frequency shift at 800 nm is* $\Delta\omega_{nl}^{silica}(800nm)=4878\ THz$.

On Fig. 3a) is presented the graphics of group-phase velocity difference for the diapason from 780 nm up to 2 µm. After that it is easily to calculate the nonlinear frequency shift by the expression $\Delta\omega_{nl}^{bk7}=3k_0(v_{ph}-v_{gr})$. On Fig. 3b) is presented the nonlinear shift for the same spectral region. The calculated nonlinear frequency shift for BK7 glass at 800 nm is

$$\Delta\omega_{nl}^{bk7}(800\ nm)=3k_0(v_{ph}-v_{gr})=48.78\ THz\ . \qquad (15)$$

This corresponds to a shift towards the short wave lengths with delay

$$\Delta\lambda=\frac{n\lambda^2}{2\pi c}\Delta\omega_{nl}=25.1\quad nm\ . \qquad (16)$$

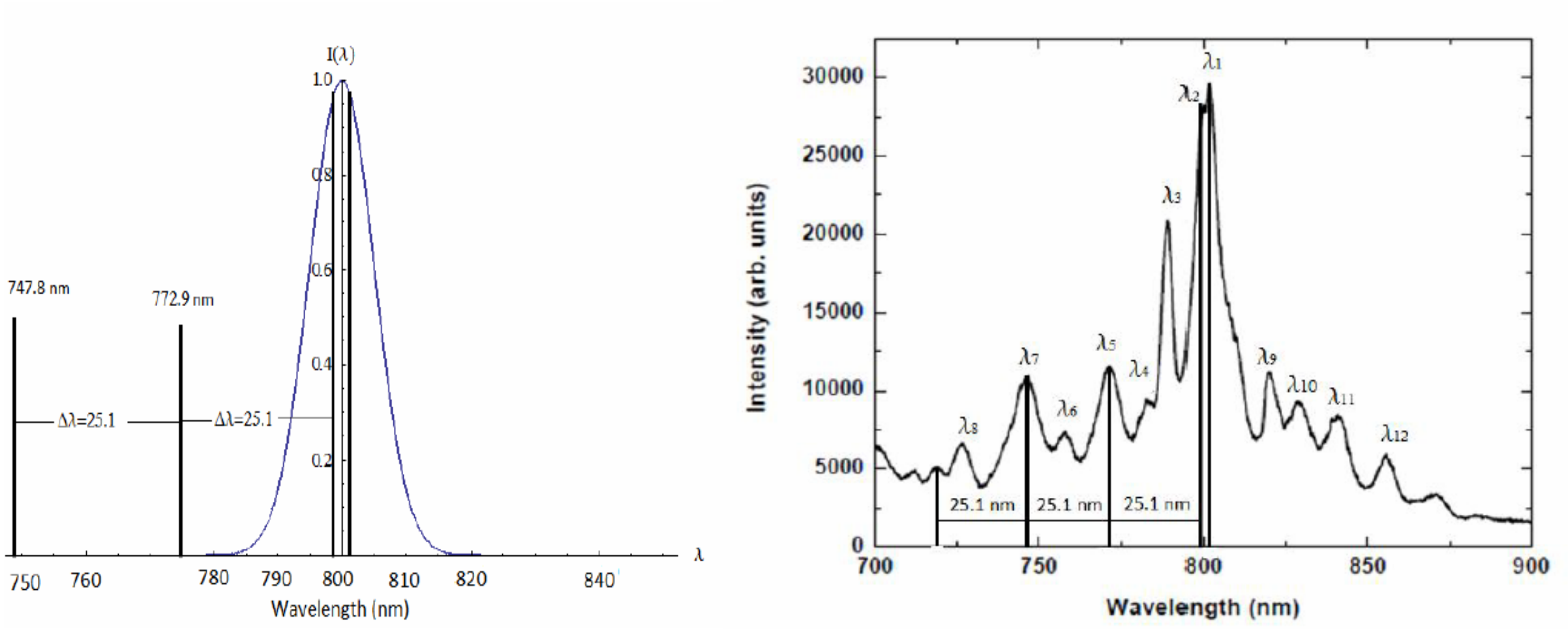

**a)** **b)**

*Fig. 4. a) Expected shift towards the short wavelengths obtained from the physical model, due to THz generation with wavelength delay* $\Delta\lambda = \frac{n\lambda^2}{2\pi c}\Delta\omega_{nl} = 25.1\ nm$ *. b) Strong spectral components with the same spectral displacement observed in the experiment.*

On Fig. 4a) is presented the expected spectral distance between the main wave and the cascaded signal waves due to nonlinear THz generation in BK7 glass with wavelength delay $\Delta\lambda = \frac{n\lambda^2}{2\pi c}\Delta\omega_{nl} = 25.1\ \ nm$. The strong spectral components on Fig. 4b) observed in the experiment are with the same spectral displacement $\Delta\lambda = 25.1\ \ nm$. There is full coincidence between the experimental measurements and the theoretical results. A careful analysis shows that the other components satisfy FPPM conditions.

*4.3 Mathematical model*

I. Three waves

The normalized system of equations in the case of three waves, including THz wave synchronism conditions $3\omega_1 = \omega_2$ and $3\omega_3 = \omega_1$ and FPP mismatch condition $\omega_3 + \omega_2 = 2\omega_1$ is

$$i\frac{\partial A_i}{\partial t} = -\beta_i\frac{\partial^2 A_i}{\partial z^2} + \gamma_i P_{nl}^i \quad i = 1\ldots3, \tag{17}$$

with polarization components

$$P_{nl}^1 = \left(|A_1|^2 + 2|A_2|^2 + 2|A_3|^2\right)A_1 + A_1^{*2}A_2\exp(i\Delta kz) + 2A_1^*A_2A_3\exp(i\Delta kz) + \frac{1}{3}A_3^3\exp(-i\Delta kz) \tag{18}$$

$$P_{nl}^2 = \left(|A_2|^2 + 2|A_1|^2 + 2|A_3|^2\right)A_2 + A_1^2A_3^*\exp(-i\Delta kz) + \frac{1}{3}A_1^3\exp(-i\Delta kz) \tag{19}$$

$$P_{nl}^3 = \left(|A_3|^2 + 2|A_1|^2 + 2|A_2|^2\right)A_3 + A_1^2A_2^*\exp(-i\Delta kz) + A_3^{*2}A_1\exp(i\Delta kz), \tag{20}$$

where $\beta_j = k_j'' v_{gr}^2 / z_0$ are the dimensionless dispersion parameters and $\gamma_j = n_2 |A_0|^2 k_j z_0$ are the nonlinear coefficients. We use localized initial conditions in the form

$$A_j = A_{0j} \exp(i\varphi_j)\exp(i\Delta\lambda_j z)\mathrm{sech}(z / z_{0j}) \quad j = 1 \cdots 3, \tag{21}$$

where $A_{0j}$ are the amplitude constants, $\varphi_j$ are the initial phases, $\Delta\lambda_j$ are the spectral shifts in respect to main wavelength, and $z_{0j}$ are the initial width of the pulses. The numerical experiments are performed by using the split-step Fourier method. On Fig. 5 is presented numerical simulation for the following values of parameters $\gamma = 0.7$, $\beta_j = 0.01$ and initial constants

$$A_{01} = 1, A_{02} = A_{03} = 0.04, \quad z_{01} = 1, \quad z_{02} = z_{03} = 5, \quad \Delta\lambda_1 = 0, \quad \Delta\lambda_2 = 2.5, \quad \Delta\lambda_3 = -2.5 \tag{22}$$

On Fig. 5a) is plotted the generation of signal waves in BK7 glass corresponding to time propagation at time 5 ps, 10 ps, 15 ps ($z = v_{gr} t$). Strong energy exchange between the carrying wave and the signal waves is observed. On Fig. 5b) is presented the spectrum of the pulses at the same time distances. Due to the THz generation we obtain asymmetrical spectrum and significant increasing of the spectral component of the signal wave towards the short wavelengths.

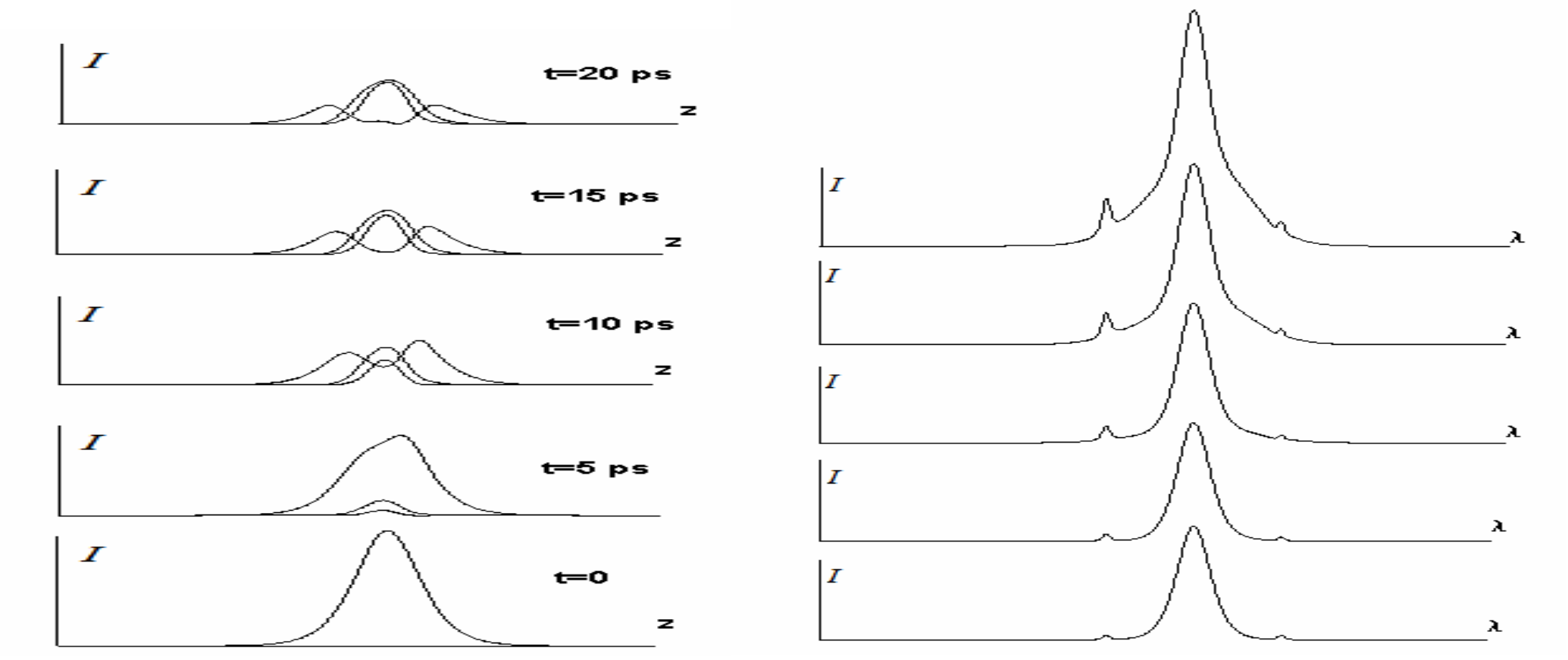

**a)** **b)**

*Fig. 5. a) Generation of signal waves at different distances in BK7 glass corresponding to time propagation t=5 ps, 10 ps, 15 ps ($z = v_{gr} t$). b) The spectrum of the pulses on the same distances. Due to the THz generation we obtain asymmetrical spectrum.*

*II. Five waves*

The system of equations in the case of five waves including THz wave synchronism conditions

$$3\omega_1 = \omega_2; \; 3\omega_3 = \omega_1 \; ; \; 3\omega_5 = \omega_3; \; 3\omega_2 = \omega_4, \tag{23}$$

and FFP mismatch conditions

$$\omega_3 + \omega_2 = 2\omega_1; \; \omega_4 + \omega_1 = 2\omega_2; \; \omega_1 + \omega_5 = 2\omega_3; \; \omega_4 + \omega_5 = 2\omega_1, \tag{24}$$

where the frequencies are situated in order from high to low frequency $\omega_4 - \omega_2 - \omega_1 - \omega_3 - \omega_5$

is

$$i\frac{\partial A_i}{\partial t} = -\beta_i \frac{\partial^2 A_i}{\partial z^2} + \gamma_i P_{nl}^i \quad i = 1\ldots 5\,. \tag{25}$$

The polarization components are

$$\begin{aligned}
&P_{nl}^1 = \left(|A_1|^2 + 2|A_2|^2 + 2|A_3|^2 + 2|A_4|^2 + 2|A_5|^2\right)A_1 + \\
&A_1^{*2}A_2\exp(i\Delta kz) + 2A_1^* A_2 A_3\exp(i\Delta kz) + \frac{1}{3}A_3^3\exp(-i\Delta kz) + 2A_1^* A_4 A_5\exp(i\Delta kz) + \\
&A_2^2 A_4^*\exp(-i\Delta kz) + A_3^2 A_5^*\exp(-i\Delta kz)
\end{aligned} \tag{26}$$

$$\begin{aligned}
&P_{nl}^2 = \left(|A_2|^2 + 2|A_1|^2 + 2|A_3|^2 + 2|A_4|^2 + 2|A_5|^2\right)A_2 + \\
&A_2^{*2}A_4\exp(i\Delta kz) + 2A_2^* A_1 A_4\exp(i\Delta kz) + \frac{1}{3}A_1^3\exp(-i\Delta kz) + A_1^2 A_3^*\exp(-i\Delta kz)
\end{aligned} \tag{27}$$

$$\begin{aligned}
&P_{nl}^3 = \left(|A_3|^2 + 2|A_1|^2 + 2|A_2|^2 + 2|A_4|^2 + 2|A_5|^2\right)A_3 + \\
&A_3^{*2}A_1\exp(i\Delta kz) + A_1^2 A_2^*\exp(-i\Delta kz) + \frac{1}{3}A_5^3\exp(-i\Delta kz) + 2A_1 A_5 A_3^*\exp(i\Delta kz)
\end{aligned} \tag{28}$$

$$\begin{aligned}
&P_{nl}^4 = \left(|A_4|^2 + 2|A_1|^2 + 2|A_2|^2 + 2|A_3|^2 + 2|A_5|^2\right)A_4 + \\
&A_1^2 A_5^*\exp(-i\Delta kz) + A_2^2 A_1^*\exp(-i\Delta kz) + \frac{1}{3}A_2^3\exp(-i\Delta kz)
\end{aligned} \tag{29}$$

$$\begin{aligned}
&P_{nl}^5 = \left(|A_5|^2 + 2|A_1|^2 + 2|A_2|^2 + 2|A_4|^2 + 2|A_4|^2\right)A_5 + \\
&A_3^2 A_1^*\exp(-i\Delta kz) + A_1^2 A_4^*\exp(-i\Delta kz) + A_5^{*2}A_3\exp(i\Delta kz)
\end{aligned}\,. \tag{30}$$

On Fig. 6 is plotted the spectrum of the pulse with five spectral components at the same time distances (t = 5 ps, 10 ps, 15 ps ($z = v_{gr}t$)). The presented numerical experiment is performed for the following values of parameters $\gamma = 0.7, \quad \beta_j = 0.01$ and initial constants

$$\begin{aligned}
&A_{01} = 1,\, A_{02} = A_{03} = 0.05, \quad A_{04} = A_{05} = 0.03 z_{01} = 1, \quad z_{02} = z_{03} = z_{04} = z_{05} = 5, \\
&\Delta\lambda_1 = 0, \quad \Delta\lambda_2 = 2.5, \quad \Delta\lambda_3 = -2.5, \quad \Delta\lambda_4 = 5, \quad \Delta\lambda_3 = 5.
\end{aligned}\,. \tag{31}$$

The asymmetry becomes stronger than in the case of three waves. Thus, by combined THz generation and FFPM processes, we obtain again asymmetrical spectrum towards the short wavelengths.

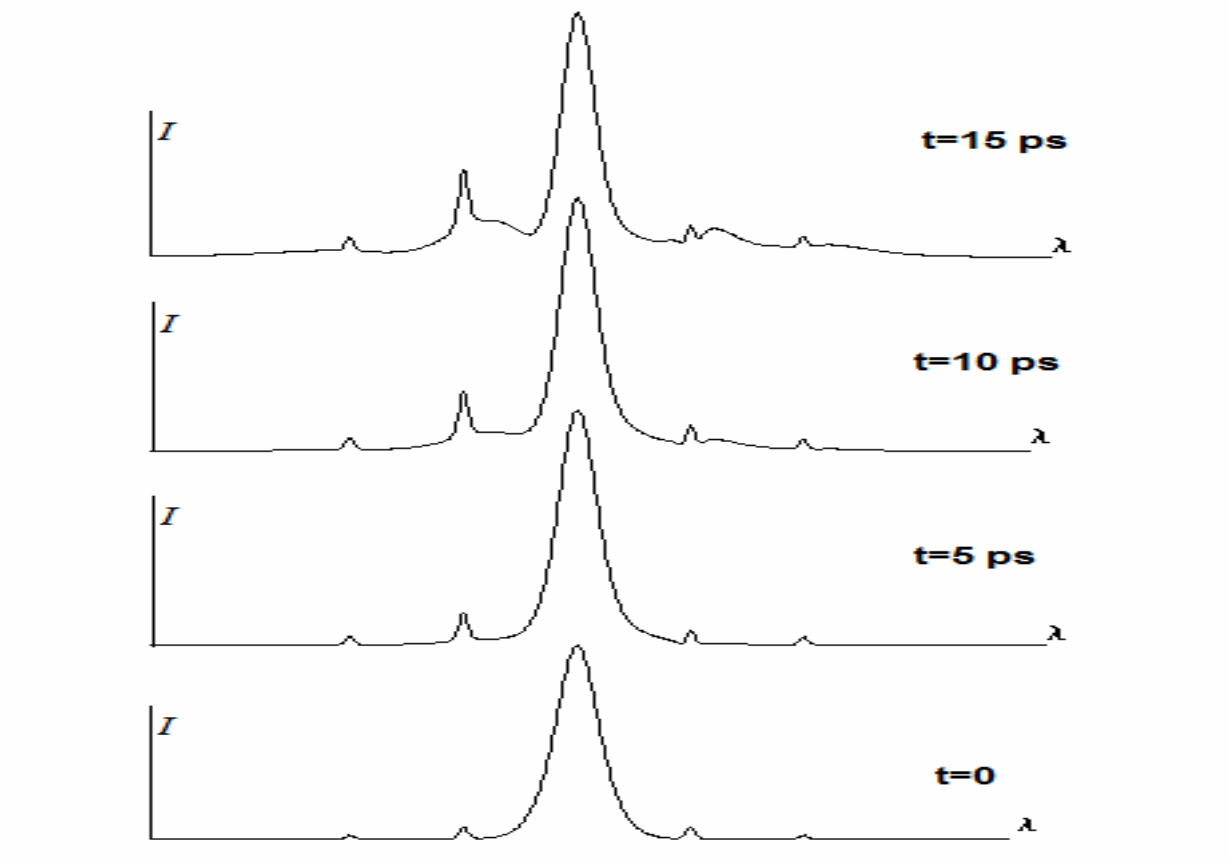

*Fig. 6. The spectrum of the pulses at time distances t=5 ps, 10 ps, 15 ps* $(z = v_{gr} t)$*. The THz generation dominates over the FPP processes and the asymmetry becomes stronger.*

## 5. Nonlinear mechanisms of merging and energy exchange between the filaments.

The recent experiments with high power Ti: Sapphire laser pulses demonstrate that it is not possible to produce a homogeneous beam pattern. Hot zones are situated across the beam cross section. Each hot zone self-focuses into a filament if the intensity and the power are high enough. Each of the multiple filaments has a core intensity clamped down to that of a single filament of the order of 0.5-5 TW/cm$^2$ [19]. These intensities are two to three orders less than the intensity needed for defocusing by ionization. The filaments with such intensities attract and exchange energy during their propagation. The final result is that at long distances survive only few of them [14, 19]. The absence of ionization in these processes forced us to seek other nonlinear mechanisms for description of the above mentioned effects. That is why in [20] we investigate two types of nonlinear interaction between collinear femtosecond laser pulses with power slightly above the critical for self-focusing $P_{cr}$. In the first case we study energy exchange between filaments. The model describes this process through a degenerate four-photon parametric mixing (FPPM) scheme and requests initial phase difference between the waves. When there is no initial phase difference between the pulses, the FPPM process does not work. In this case the second type of interaction is obtained as merging between two, three or four filaments in a single filament with higher power. It is found that in the second case the interflow between the filaments has a potential of interaction due to cross-phase modulation (CPM) [20].

Usually, the phases between the optical pulses are not fixed and in the processes of interaction take place both CPM and FPPM. On Fig. 7 is presented numerical experiment, when between the pulses there is initial phase difference. As a result of the interaction both attraction due to CPM and energy exchange due to FPPM are obtained. We use the fact, that the optical pulses $\vec{A}_j$ admit both polarization components. Let us represent each component as $\vec{A}_j = \vec{A}_{jx}\vec{x} + A_{jy}\vec{y}$, $j = 1,2\cdots n$. Then the initial conditions for investigation of interaction of three filaments in the frame of VSNSE (13) after writing the system in dimensionless coordinates are

$$A_x = \sum_{j=1}^{3} A_{jx} = \frac{A_x^{01}}{\sqrt{2}} \exp\left( -\frac{(x+a)^2 + y^2 + z^2}{2} \right) \exp(i\varphi_1) + \frac{A_x^{02}}{\sqrt{2}} \exp\left( -\frac{(x-a)^2 + y^2 + z^2}{2} \right) \exp(i\varphi_2)$$

$$+ \frac{A_x^{03}}{\sqrt{2}} \exp\left( -\frac{x^2 + (y-b)^2 + z^2}{2} \right) \exp(i\varphi_3)$$

$$A_y = \sum_{j=1}^{3} A_{jy} = \frac{A_y^{01}}{\sqrt{2}} \exp\left( -\frac{(x+a)^2 + y^2 + z^2}{2} \right) \exp(i\varphi_1) + \frac{A_y^{02}}{\sqrt{2}} \exp\left( -\frac{(x-a)^2 + y^2 + z^2}{2} \right) \exp(i\varphi_2) \text{ ,(32)}$$

$$+ \frac{A_y^{03}}{\sqrt{2}} \exp\left( -\frac{x^2 + (y-b)^2 + z^2}{2} \right) \exp(i\varphi_3)$$

where $A_x$ and $A_y$ are composition of *x*- and *y*-components of three optical pulses situated in different places in the x-y plane. The presented numerical simulation on Fig. 7 is carried out for $a = 3.2$, $b = 2$ (*a* and *b* are the displacements of the centum of weight of each component), $\varphi_1 = 0$, $\varphi_2 = \pi / 2$, $\varphi_3 = 0$ (the initial phase differences between the pulses).

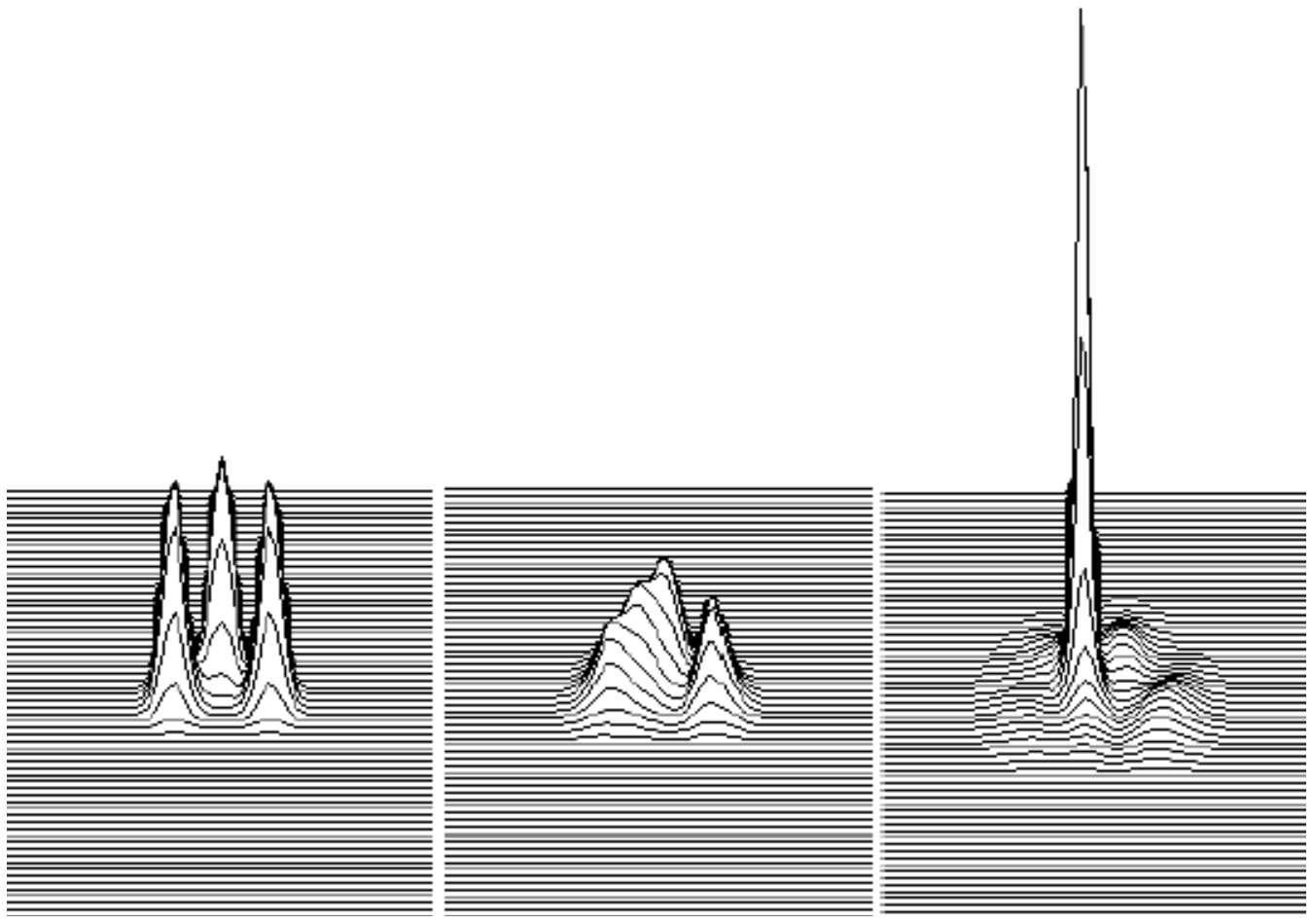

**z = 0** **z = 5/2 $z_{diff}$** **z = 5 $z_{diff}$**

*Fig 7. Evolution of three Gaussian laser pulses propagating at 5 diffraction lengths and governed by normalized VSNAE (13) under initial conditions (32). The intensity is slightly above the critical for self-focusing. An attraction due to CPM and energy exchange due to FPPM can be observed. Projection of the spots of the pulses* $(x, y)$ *is plotted.*

The numerical experiments with higher number of initial pulses with arbitrary phases present the general trend observed in the experiments – reduction of the number of pulses, merging, energy exchange and self-compression.

## 6. Conclusions

The experiments with high power fs laser pulses demonstrate the possibility for generation of stable filaments in gases and silica glasses without ionization of the media. In air the intensity region, where the filament exists is in the range $10^{11} - 10^{12}$ $W / cm^2$, which is slightly above the critical for self-focusing. These intensities are two - three orders lower than the intensity needed for defocusing by ionization. The observation of filaments and white continium in silica glasess is also in regime without plasma channels. The absence of ionization in these processes forced us to look for other nonlinear mechanisms for description of the above mentioned effects. We present new parametric conversion mechanism for asymmetric spectrum broadening of femtosecond laser pulses towards the higher frequencies in isotropic media. This mechanism includes cascade generation with THz spectral shift for solids and GHz spectral shift for gases. This shift is equal to 3CEP $\omega_{nl} = 3k_0(v_{ph} - v_{gr})$. The process works simultaneously with the four-photon parametric wave mixing. The proposed theoretical model gives very good coincidence with the experimental data. In addition we demonstrate that the presented nonlinear model describes the process of reduction of the number of filaments by two mechanisms of nonlinear interaction: attraction due to CPM and energy exchange due to FPPM.

## 7. Acknowledgements

The authors acknowledge the support form Bulgarian National Science Fund by grant DFNI–I-02/9 and also under the project "Laser induced fabrication of three dimensional nanoparticles structures and study of their optical properties”.